# An LLM-RAG Approach for Healthy Eating Index-Informed Personalized Food Recommendations

**Yibin Wang[1], Yanjie Yang[2], Grace Melo Guerrero[2], Rodolfo M. Nayga Jr. [2], and Azlan Zahid[1, *]**
[1] Department of Biological and Agricultural Engineering, Texas A&M AgriLife Research, Dallas, TX, USA
[2] Department of Agricultural Economics, Texas A&M University, College Station, TX, USA
*Corresponding author: azlan.zahid@tamu.edu

**Abstract**

Diet quality is a leading determinant of chronic disease risk. Advances in artificial intelligence (AI) have enabled food recommendation systems to adapt suggestions to user preferences and health goals. However, most current systems rely on loosely curated food databases and provide limited connection to a validated index. In this study, we propose a Healthy Eating Index (HEI)-informed retrieval-augmented generation (RAG) framework that combines standardized nutrition databases with large language models (LLMs) for personalized food recommendations. Our proposed method anchors retrieval in the National Health and Nutrition Examination Survey (NHANES) and the Food Patterns Equivalents Database (FPED). A food-level embedding space is constructed from FPED-derived textual descriptions. For each entity, the system computes baseline HEI scores, retrieves candidate foods for intake recommendations, and estimates the HEI impact of simple substitutions or additions. A constrained RAG pipeline instantiated with a pretrained OpenAI LLM generates personalized recommendations and sources based on nutrient profiles and HEI contributions. The simulation results showed a mean HEI improvement of 6.45 ± 4.02, with the proportion of users' HEI over 50 increasing from 45.12% to 61.26%. Quantile analysis revealed consistent improved shifts across the HEI distribution. Our findings suggest that the proposed LLM-RAG-based AI systems can support more precise, explainable, and personalized nutrition guidance to improve diet quality.

## 1. Introduction

Diet quality is a critical, yet modifiable, factor in overall human health. However, most adults struggle to translate nutrition science into simple, daily choices that accumulate into better long-term outcomes (Silva et al., 2023). Beyond knowledge gaps, day-to-day constraints such as time pressure, food access, food cost, cultural preferences, and taste preferences shape what is feasible at the point of dietary decision (Cowan-Pyle et al., 2024). Consequently, generic dietary guidance seldom translates into actionable choices without context-aware personalized recommendations. In addition, dietary assessment is a central bottleneck in this effort. Conventional dietary assessment tools, such as 24-hour recalls and food frequency questionnaires, are well-validated and widely used in research and surveillance (Subar et al., 2012), yet they appear resource-intensive, cognitively demanding, and difficult to integrate into continuous self-management. These approaches are typically administered at discrete time points and are not designed to deliver real-time feedback or adapt to an individual's dietary patterns and rather provide recommendations for the average individual. As a result, estimated intakes of nutrients that are critical for chronic disease prevention and diet quality can be systematically biased, limiting the extent to which traditional assessment methods can directly support personalized dietary recommendations.

Decades of public health investment have produced robust population-level nutrition infrastructures: the National Health and Nutrition Examination Survey (NHANES) for nationally representative intake distributions (Akinbam et al., 2022), the Food Patterns Equivalents Database (FPED) to map reported foods to dietary components (Bowman et al., 2020), and the Healthy Eating Index (HEI) as a validated measure of overall diet quality (Krebs-Smith et al., 2018). Nevertheless, these assets are still primarily used for surveillance and research, and are rarely operationalized jointly to deliver individualized, explainable recommendations.

The HEI has emerged as a widely used metric of how closely an individual's intake aligns with the dietary guidelines. Once HEI scores and component profiles are computed from detailed intake data linked to FPED, they provide a rich summary of strengths and gaps across food groups and nutrients (Kirkpatrick et al., 2018). However, this information remains largely descriptive, where HEI alone does not specify how a given individual might adjust their habitual eating pattern in feasible ways. Existing commercial diet tools tend to focus on calories and macronutrients, rely on crowdsourced food databases with heterogeneous provenance and limited quality control, and offer generic, rule-based tips that fail to explicitly target HEI components or leverage research-grade intake data (Zhang et al., 2025). This gap motivates the need for algorithmically translating HEI-defined deficits and population-based reference data into real-time personalized food recommendations firmly grounded in established nutrition science. Bridging this gap can have a substantial public health impact by enabling more precise dietary guidance at the individual level, potentially improving diet quality and reducing risk for chronic diseases. From a policy perspective, operationalizing these into actionable recommendations could inform evidence-based interventions and enhance the effectiveness of nutrition programs and dietary guidelines.

Recent advances in artificial intelligence (AI) and large language models (LLMs) have emerged as promising foundations for smart food recommendation methodologies (Adhikari & McFadden, 2025; Bondevik et al., 2024; Tsolakidis et al., 2024). Large-scale dietary datasets coupled with nutrient and food pattern databases can be employed to construct rich vector representations, i.e., embeddings of foods and meals based on nutritional properties (Bergling et al., 2025; Sharma et al., 2025). The LLMs can provide reasoning and natural language generation, interpretable corpus of foods, nutrient profiles, and guideline-consistent patterns (Singh et al., 2024). Popular architectures such as GPT, Llama, and DeepSeek have been explored for nutrition counseling and food recommendation, revealing that LLMs can generate accurate suggestions and structured meal plans (Khamesian et al., 2025). Deep generative networks were adopted to produce personalized meal plans that are highly accurate in both energy intake and nutritional content, by incorporating user-specific information and enforcing well-defined nutritional rules (Papastratis et al., 2024).

Researchers have also explored hybrid strategies that combine dense and sparse text embeddings to improve recall of relevant items, as well as LLM-based systems for nutrition suggestions in conversational settings (Zhang et al., 2025). Building on advances of LLMs, an emerging research stream emphasizes personalization and explainability by integrating personal and population models (Yang et al., 2024). Transformer-based methods have also demonstrated a strong performance in tasks such as ingredient understanding, recipe retrieval, and preference-aware ranking (Mereddy & Beedareddy, 2024; Rani et al., 2024). However, existing studies relied on loosely curated databases and lacked explicit grounding in standardized population reference datasets and formal diet quality indices such as the HEI. As a result, the recommendations generated are difficult to compare across populations and challenging to interpret in terms of established dietary guidelines.

To incorporate LLMs with trustworthy nutrition knowledge, Retrieval-Augmented Generation (RAG) has emerged as a promising approach that grounds model outputs in structured external retrieved evidence (Lewis et al., 2020). RAG frameworks can connect LLMs to structured knowledge sources, allowing agents to generate context-aware recommendations grounded in evidence-based dietary patterns rather than generic internet textual data. Recently, a RAG system was developed to deliver personalized fitness and dietary guidance by integrating real-time wearable data with curated knowledge sources (Pyae et al., 2025). By following evidence-based nutritional guidelines and emphasizing eco-friendly ingredients, the RAG-embedded web application enabled individuals with obesity and type 2 diabetes to make informed, sustainable dietary choices that align with recommended management strategies (Gavai & Van Hillegersberg, 2025). The effectiveness of RAG-enhanced LLMs was also evaluated in delivering guideline-adherent nutrition information for cardiovascular disease prevention (Parameswaran et al., 2025). RAG pipelines were investigated in conversational AI to generate contextualized responses by integrating the user query and preferences with retrieved recipes, statistical summaries, and substitution logic (Tsampos & Marakakis, 2025). Furthermore, graph-based RAG algorithms have been proposed for representation learning on heterogeneous graph-structured data, using attention mechanisms to reconstruct node features and edges while employing a dual hierarchical attention scheme (Forouzandeh et al., 2024; S et al., 2025).

Nevertheless, methodological gaps remain in existing RAG-based food recommendation studies regarding HEI-informed precision nutrition applications. Most current systems do not explicitly encode HEI component scores or population-reference intake distributions, and LLMs are less constrained in their generative flexibility. This can increase the risk of recommendations that drift from established nutrition science or overlook relevant diet-quality targets. These limitations motivate the development of HEI-aware RAG-based methodologies that systematically integrate standardized population data and individual profiles to produce transparent and trustworthy personalized dietary recommendations.

In this study, we propose a HEI-aware RAG-LLM framework for personalized food recommendation anchored in authoritative nutrition databases. The objective is to generate daily meal suggestions that particularly target HEI component deficits while preserving interpretable sources. Within the proposed framework, individual health conditions are first estimated from user profiles and recent intake summaries. The system then retrieves candidate foods that are expected to improve HEI components, subject to user-specific health profile. Generated recommendations report anticipated healthy eating contributions and offer multiple portion variants to support gradual behavior change. The HEI-aware retrieval module indexes foods in an embedded nutrient-pattern space and identifies candidate substitutions or additions. The RAG pipeline then connects a pretrained LLM to these structured knowledge sources, constraining it to cite retrieved nutrient entries from standardized food databases rather than relying on generic internet text. Rather than generating recommendations as a generic similarity search, we define HEI-informed retrieval objectives that prioritize foods expected to improve the dietary components. This design allows the system to generate recommendations that are interpretable in terms of HEI and accessible to researchers and practitioners for evaluating the intended aspects of diet quality. In addition, the system remains compatible with established dietary assessment workflows and can be extended or recalibrated in line with guidelines. The main contributions of this study are summarized as:

1. Development of a HEI-informed RAG-LLM framework for personalized food recommendation anchored in user health profile and structured food databases.
2. Design of a retrieval mechanism that restricts pretrained LLM to operate over retrieved foods and structured nutrient metadata, generating interpretable personalized recommendations with anticipated HEI contributions.
3. Evaluation of the proposed framework on 12,076 user records, demonstrating an average HEI improvement of $6.45 \pm 4.02$ and an increase in the proportion of individuals with HEI > 50 from 45.12% to 61.26%.
4. This study advances personalized nutrition research by translating HEI metrics into actionable recommendations and demonstrates a scalable tool that could directly inform public health programs and dietary policy.

## 2. Materials and Methods

### 2.1 Data Acquisition

The datasets used in this study were obtained from the publicly available National Institutes of Health (NIH) NHANES 2017 – March 2020 Pre-pandemic cycles and United States Department of Agriculture (USDA) FPED (Akinbam et al., 2022; Bowman et al., 2020). NHANES is a nationally representative continuous survey that combines in-home interviews, standardized physical examinations, and 24-hour dietary recalls. We utilized the total nutrient intake (DRTOT) files from the dietary recall component. These data provide total daily intake of energy, macronutrients, micronutrients, and selected dietary components for each respondent derived from 24-hour recalls. From DRTOT, we extracted variables including total energy, protein, carbohydrate, total fat, dietary fiber, sodium, potassium, and selected vitamins and minerals. When multiple recall days were available for a respondent, intakes were averaged across days to obtain a more stable estimate of usual daily intake. The FPED was used to map reported intake to standardized food groups and limit components. FPED population-level files were linked to the NHANES total nutrient intake files (DRTOT) using respondent identifiers to derive food pattern equivalents at the person-day level. After linking to FPED, we aggregated these food pattern equivalents at the person-day level to obtain a set of food groups and limit components for each respondent and recall day. These FPED-derived variables serve as the basis for computing HEI-2020 diet quality indicators and constructing HEI-relevant features used in our recommendation algorithm.

In addition to dietary intake, we obtained NHANES demographic data (DEMO), body measures (BMX), diabetes (DIQ), and diet behavior (DBQ) data to define user profile for each respondent. Specifically, we extracted demographic characteristics including age, sex, race/ethnicity, education, and household income indicators, as well as health status indicators such as self-reported or clinically defined diabetes, cardiovascular disease, and iron-related measures. These variables were obtained to capture both sociodemographic context and diet-sensitive clinical conditions that are relevant for generating recommendations. The inputs were then aggregated to approximate the health profile available for a real user in a personalized nutrition setting. In total, 15,560 users with 383 baseline features were extracted from the public databases prior to data preprocessing, irrespective of missing or incomplete values. All records were linked across NHANES and FPED data files using the unique respondent identifier (SEQN) and corresponding food codes to ensure consistent integration of dietary, demographic, and health-related information.

## 2.2 Data Preprocessing

We first concatenated all relevant NHANES 2017 – March 2020 Pre-pandemic data to construct a unified person-level dataset. DEMO, BMX, DIQ, DBQ, and day-1 and day-2 total nutrient intake files (DR1TOT, DR2TOT) were merged using the unique identifier SEQN. The merged records were then linked to FPED-derived variables via NHANES food codes. We then applied a simple quality filter based on HEI availability. After merging user profiles and dietary intake files, the combined dataset was linked to the HEI-2020 output table using SEQN. Only respondents with a valid HEI-2020 score were retained. Respondents with missing values in core demographic variables, including age, sex, race/ethnicity, anthropometrics such as BMI, key health indicators including diabetes, cardiovascular disease, or essential nutrients and FPED components were excluded. We also removed records lacking valid dietary recalls or HEI scores. After this filtering process, the analytic dataset consisted of 12,076 respondents and 388 variables including HEI scores, representing a complete cross-linked profile for each individual. HEI component and total scores were then computed according to the official 2020 USDA scoring algorithm applied to the NHANE-FPED population-level data (Shams-White et al., 2023). For respondents with two valid recall days, HEI scores were averaged across days to approximate usual diet quality. On the other hand, for respondents with a single valid recall, day-1 scores were retained. The resulting HEI-2020 total and component scores were merged back into the person-level dataset and used to define baseline diet quality and component-specific anchors.

In parallel, we prepared a food-level corpus for retrieval and embedding. FPED records were aggregated by food code, and each food item was converted into a structured text description summarizing its nonzero FPED components, for instance, cup or ounce equivalents of major food groups and components. These descriptions and their associated metadata, including food codes and short descriptions, were wrapped as documents and encoded into dense embeddings using a sentence Transformer model MiniLM-L6-v2 (Wang et al., 2020). The resulting vectors were stored in an FAISS index (Douze et al., 2025), together with a JSONL file containing the corresponding texts and metadata, forming the nutrition-aware embedding space from which the RAG-LLM retrieves candidate foods in the subsequent modeling stage.

After feature construction, we derived two linked analytic datasets. The person-level dataset contains one record per respondent, including user profile variables, disease indicators, preference-related features, and summary dietary variables, including HEI-2020 total and component scores. The food-level dataset consists of FPED-mapped food items with their corresponding nutrient profiles and food pattern equivalents, prepared for embedding and retrieval within the proposed RAG-based recommendation framework. These processed datasets served as the foundation for the model described in Section 2.3, where we construct nutrition and HEI informed embeddings and develop the retrieval-augmented recommendation algorithm.

## 2.3 Proposed RAG-LLM Recommendation System

The proposed method is a RAG framework that integrates HEI-guided retrieval over a structured food embedding space with a LLM to synthesize personalized food recommendations. The framework consists of two main components: (1) an HEI-informed retrieval and ranking module that identifies candidate foods based on their projected impact on HEI components and user profiles, and (2) a retrieval-augmented generation module that produces personalized recommendations and explanations grounded in the retrieved nutrient and FPED information. **Figure 1** illustrates the overall architecture of the proposed framework.

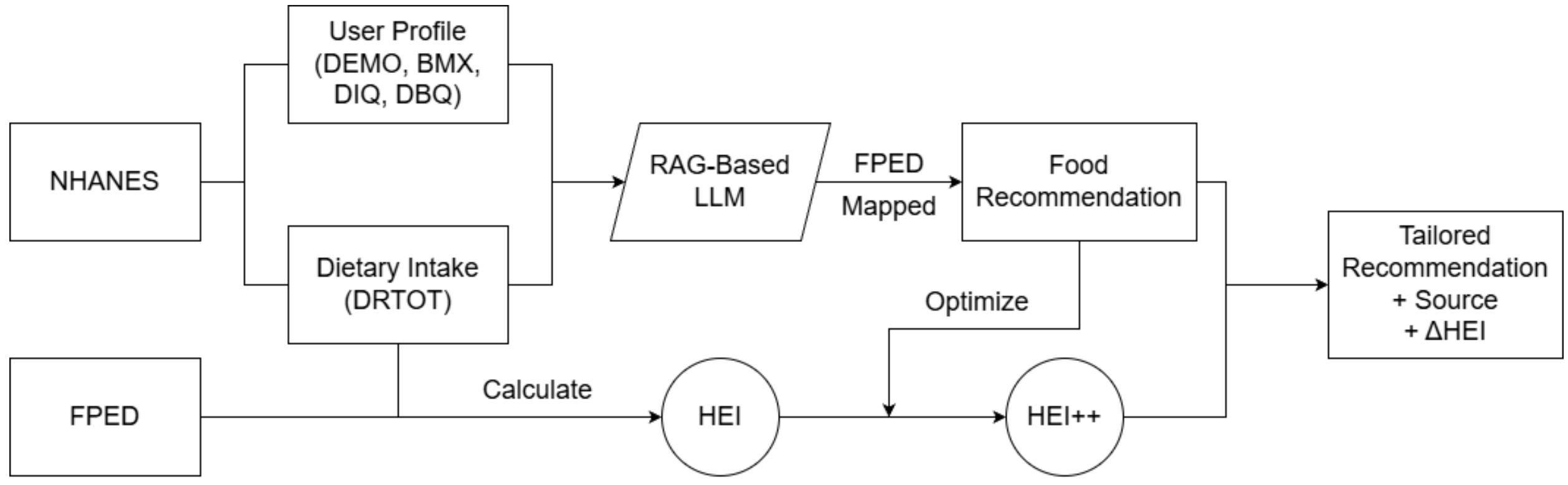

NIH National Health and Nutrition Examination Survey (NHANES); Total Nutrient Intakes (DRTOT); Demographic Data (DEMO); Body Measures (BMX); Diabetes (DIQ); Diet Behavior (DBQ); Food Patterns Equivalents Database (FPED); Healthy Eating Index (HEI); Retrieval-Augmented Generation (RAG); Large Language Model (LLM)

**Figure 1.** Flow diagram of the proposed RAG-based LLM food recommendation framework.

For the food embedding and retrieval space, let $i$ index users and $j$ index food items. Each FPED mapped food item $j$ is first converted into a structured text description summarizing its main food pattern components. These descriptions are then embedded using the sentence Transformer model to obtain a dense vector representation as $\mathbf{e}_j \in \mathbb{R}^D$, where $D$ refers to the embedding dimension. All $\mathbf{e}_j$ vectors are L2-normalized and stored in a FAISS index to support efficient similarity search.

Given a user $i$, we construct a query representation $\mathbf{q}_i \in \mathbb{R}^D$ by encoding a profile that summarizes key health and dietary information. Cosine similarity is used to retrieve foods for which the embeddings are most relevant to this profile by

$$\mathrm{sim}(\mathbf{e}_j, \mathbf{q}_i) = \frac{\mathbf{q}_i{}^{\mathrm{T}}\mathbf{e}_j}{\|\mathbf{q}_i\|_2 \|\mathbf{e}_j\|_2}. \tag{1}$$

The top K items according to $\mathrm{sim}(\mathbf{e}_j, \mathbf{q}_i)$ ranked using maximal marginal relevance from an initial candidate set $C_i$ for user $i$. We further denote the user baseline dietary intake vector by $\mathbf{x}_i$ and the corresponding HEI total score by

$$H_i = H(\mathbf{x}_i)\ , \tag{2}$$

where $H(\cdot)$ is the HEI-2020 scoring function applied to the NHANES-FPED variables. For each candidate food $j \in C_i$, we formulate a simple hypothetical modification of the user's intake $\mathbf{x}_i{}^{(j)}$ to represent the updated user intake. The expected change in HEI can be computed as

$$\Delta H_{i,j} = H\left(\mathbf{x}_i{}^{(j)}\right) - H(\mathbf{x}_i)\ . \tag{3}$$

The component-level changes $H_k\left(\mathbf{x}_i{}^{(j)}\right) - H_k(\mathbf{x}_i)$ can be tracked for selected components $k$, such as whole grains, sodium, and added sugars. Thus, the objective function that combines HEI improvement and user profile alignment with constraints can be represented as

$$J_{i,j} = \alpha \Delta H_{i,j} + \beta C_{i,j}\ , \tag{4}$$

where $\alpha, \beta > 0$ are parameters, and candidates are ranked by $J_{i,j}$ with the top K foods

$$R_i = \{j_1, j_2, \dots, j_K\}\ . \tag{5}$$

The recommended foods are selected as the final HEI-informed candidate set for the user $i$. For each selected food, the portion variant can also be computed with the corresponding $\Delta H_{i,j}$ to support graded behavior change and personal preference. The following component is the RAG-based LLM to convert the scored candidates $R_i$ into user-facing recommendations with optional source reference and concise explanations. For each user $i$, the LLM receives a structured prompt containing user profile information, current dietary pattern, and the retrieved foods $j$ with their textual descriptions, FPED components, and nutrients.

In our implementation, the LLM component is instantiated using the OpenAI ChatGPT family of models (GPT-class) (OpenAI et al., 2024), configured with a low sampling temperature to prioritize consistency and factual alignment with the retrieved nutrient data. The LLM is instructed to recommend food combinations only from $R_i$ while justifying each recommendation in terms of HEI-related improvements and health constraints. Since the model is grounded in a small, well-characterized set of retrieved candidates with known nutrient and HEI properties, the generated advice remains transparent and explainable. For instance, the system can suggest replacing a high-sodium entree with a nutritionally similar but lower-sodium option or adding a fruit and whole-grain side that jointly increases total fruit and whole-grain HEI component. This HEI-informed RAG design allows the framework to provide personalized recommendations that are consistent with established dietary guidelines.

### 2.4 Performance Evaluation

We evaluated the proposed HEI-informed RAG-LLM framework using an offline simulation on the analytic NHANES-FPED dataset containing 12,076 respondents with complete health and dietary profiles. To avoid optimistic bias in evaluating the proposed framework, we randomly partitioned the dataset into a training set and a held-out test set with a split of 80%/20%. The development set was used for exploratory analysis and selecting utility function parameters, while the test respondents were reserved as a held-out test set for reporting performance metrics.

For each respondent, the baseline HEI-2020 total score was computed from observed intake. The proposed model then generated a set of recommended foods and substitutions/additions, and we constructed a hypothetical post-recommendation intake profile to obtain a corresponding HEI score. This setup quantifies the potential diet quality gain if users fully adopt the recommended changes, without requiring real-world diet behavior data. The primary evaluation metric was the change in HEI-2020 total score at the individual level,

$$\Delta H_i = H_i^{rec} - H_i^{base} \,. \tag{6}$$

We further summarized $\Delta H_i$ across all users using mean, standard deviation, and visualized the difference between pre-recommendation and post-recommendation. In addition, we examined the proportion of users exceeding a moderate diet quality threshold (HEI > 50) before and after simulation. For a given threshold $\tau$, the proportion can be calculated as

$$p(\tau) = \frac{1}{N}\sum_{i=1}^{N} 1\{H_i > \tau\}\,, \tag{7}$$

where $1 \cdot \{\}$ is the indicator function and $N$ is the total number of users. We considered secondary metrics to further characterize the impact and plausibility of the recommendations, such as distributional shift in HEI. Changes in median, 25%, and 75% percentiles were reported for the HEI-2020 distribution to assess whether improvements were concentrated among low-HEI or more broadly distributed.

### 2.5 Prototype Implementation

To demonstrate the translational potential of the proposed LLM-RAG framework, we implemented a cross-platform mobile application using Flutter (Google LLC, Mountain View, CA, USA), deployable on both Android and iOS devices. The app communicates with a cloud-hosted backend via Firebase (Google LLC, Mountain View, CA, USA) APIs, transmitting user profile information and a brief dietary summary to the HEI-informed LLM-RAG service for processing. At the backend, the system computes baseline HEI-2020 scores, retrieves candidate foods from the FPED-based embedding index, and generates recommendations and explanations using the constrained LLM. The mobile interface then presents portion-aware food suggestions. Users can view multiple options for each eating occasion. The underlying rationale is optional to better understand how each recommendation relates to diet quality. The interface also allows users to view alternative suggestions, enabling the system to provide preferred food combinations with flexibility. In addition, the app provides a compact summary view of the user's current choice and general health status, reinforcing how individual choices accumulate into overall diet quality. A screenshot of the prototype mobile application is shown in **Figure 2** to illustrate the end-to-end user experience.

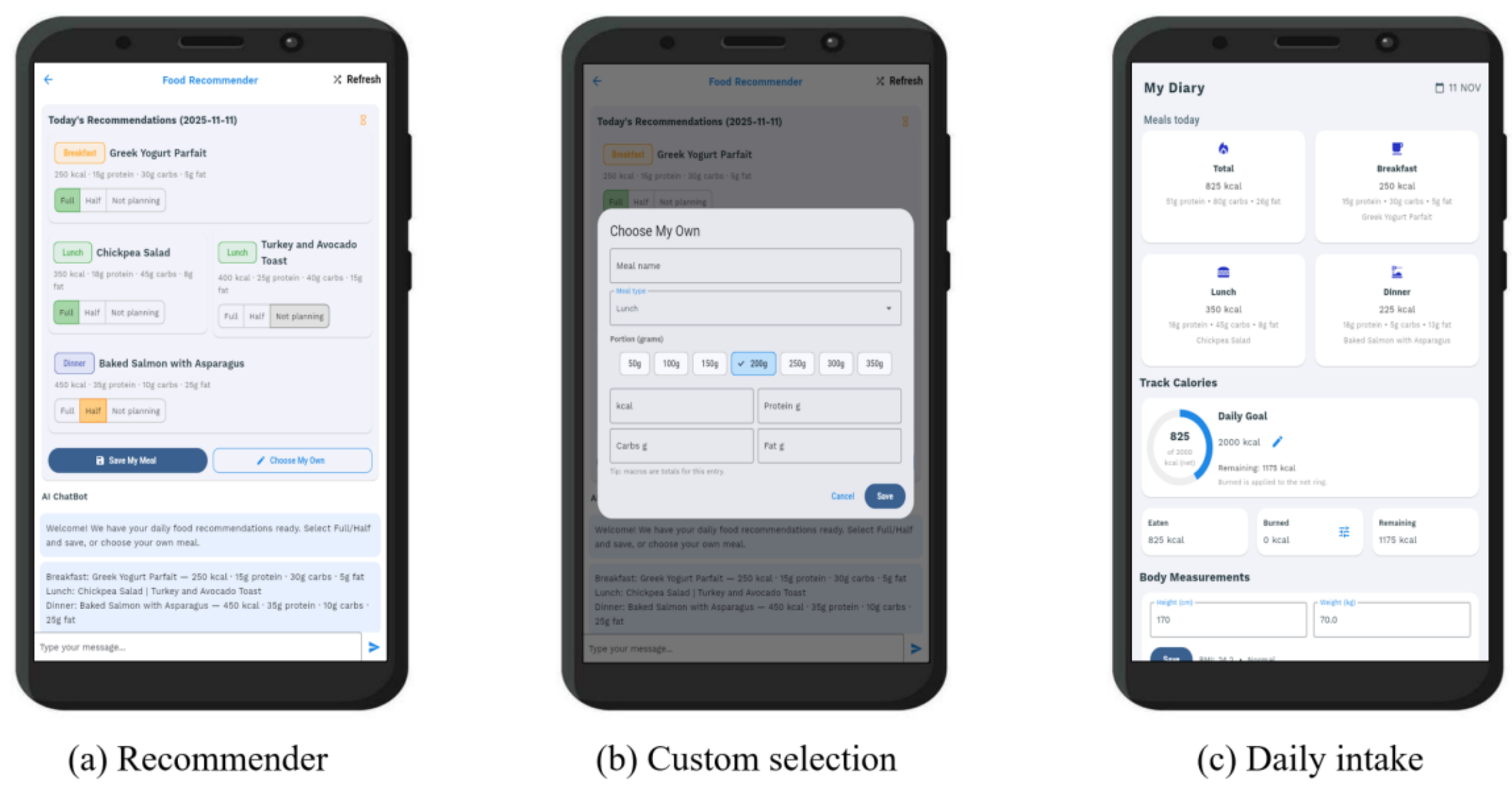

(a) Recommender (b) Custom selection (c) Daily intake

**Figure 2.** Prototype mobile application implementing the HEI-informed LLM-RAG framework for personalized food recommendations.

## 3 Results and Discussion

Simulation using the proposed HEI-aware RAG-LLM framework indicates that personalized AI-generated advice has the potential to substantially improve diet quality at the population level. The offline experiment was conducted on 2,416 test participants from the analytic databases. For each respondent, we compared the baseline HEI-2020 total score with the simulated post-recommendation HEI score derived from the modified intake profiles. Across all respondents, the mean HEI-2020 score increased by 6.45 ± 4.02 points after applying the recommended substitutions and additions, as shown in **Figure 3**. This shift corresponds to a meaningful improvement in overall diet quality, given that the HEI scale ranges from 0 to 100 and that even moderate changes in HEI have been associated with differences in cardiometabolic risk in previous cohort studies (Brauer et al., 2021). The pre-intervention and post-intervention distributions show a rightward shift, with the post-recommendation distribution more concentrated in the moderate-to-higher HEI range.

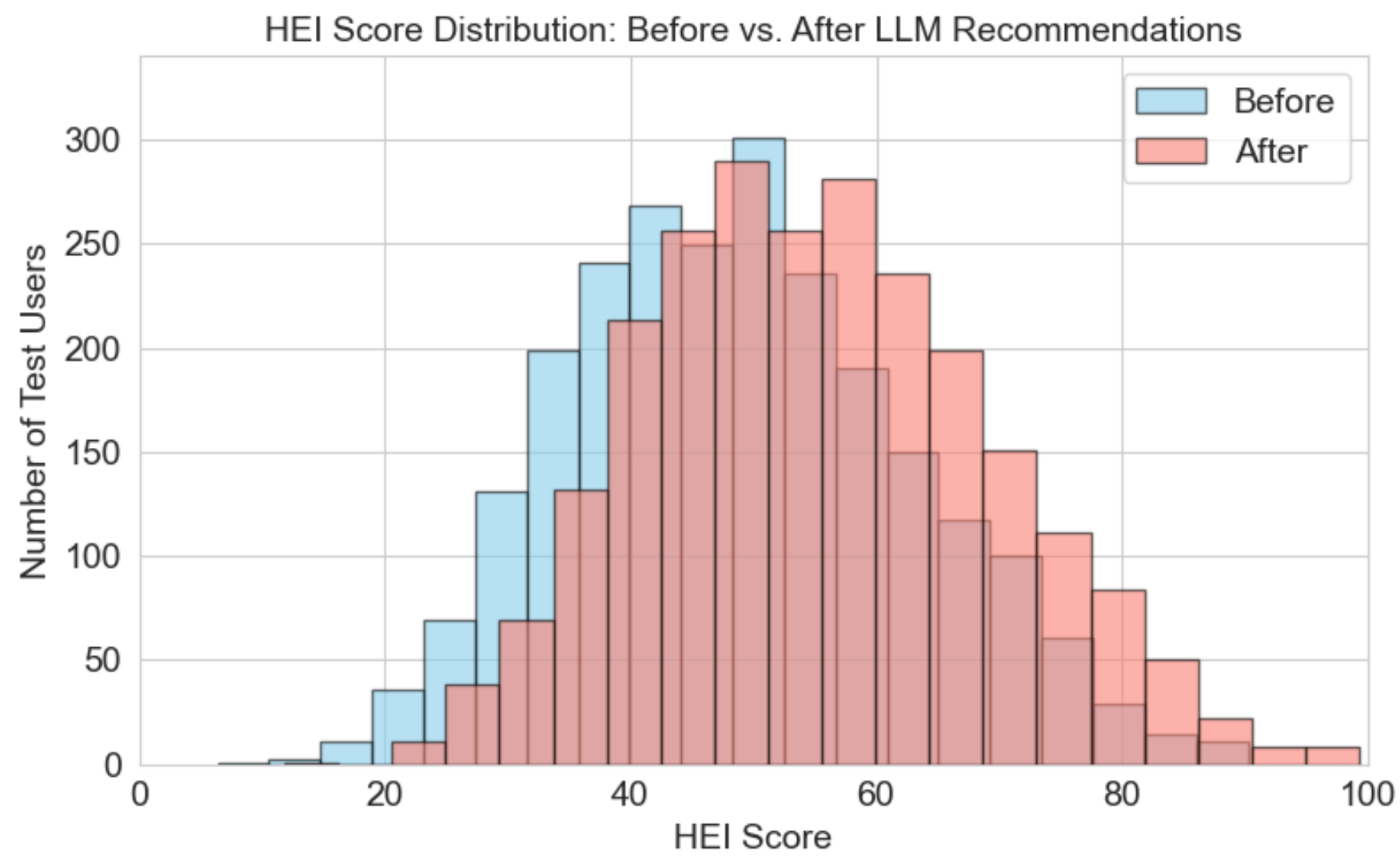

**Figure 3.** Distribution of HEI scores before and after simulated RAG-LLM recommendations.

In addition to the mean change, we examined how the recommendations affected the proportion of individuals exceeding a commonly used threshold for moderate or better diet quality. At baseline, 45.12% of users had HEI-2020 scores greater than 50. Under the simulated post-recommendation intake, this proportion increased to 61.26%, representing an increase of more than 16%. This suggests that the proposed framework not only improves average HEI but also shifts a substantial fraction of the population from lower to more favorable diet quality categories.

Moreover, the distribution of HEI scores shifted consistently toward higher diet quality. **Table 1** demonstrates that the 25th percentile increased from 38.49 to 44.84, the median from 48.32 to 54.77, and the 75th percentile from 58.07 to 64.66. These results indicate that the proposed framework improves diet quality across the entire distribution, not only among users who already had relatively high HEI scores. The absolute gains at the 25th, 50th, and 75th percentiles are of similar magnitude, suggesting that the simulated recommendations benefit both lower and higher HEI individuals in a broadly comparable way, rather than widening disparities in diet quality.

**Table 1.** HEI-2020 score quantiles before and after simulated recommendations in testing.

| Quantile | HEI Score Before | HEI Score After | ΔHEI |
|---|---|---|---|
| 25$^{th}$ Percentile | 38.49 | 44.84 | 6.35 |
| 50$^{th}$ Percentile | 48.32 | 54.77 | 6.45 |
| 75$^{th}$ Percentile | 58.07 | 64.66 | 6.59 |

The reported improvements are particularly notable in the context of vulnerable groups. Prior work has shown that low-income households typically score in the mid-50s on the HEI and are disproportionately affected by diet-related chronic conditions (Dhakal & Khadka, 2021). Within this context, an average increase of approximately 6.5 HEI points due to personalized AI-enabled advice would represent a nontrivial improvement, moving diets closer to patterns that are consistently associated with reduced risk of cardiovascular disease and other chronic conditions. Although the present analysis does not stratify results by income group, the magnitude of the simulated HEI gain at the population level is comparable to or larger than differences often observed between sociodemographic strata in observational studies.

In generated sources, the LLM consistently links recommendations to HEI-relevant dimensions, for example, highlighting that a suggested food would improve whole grain, total fruit, or sodium components while maintaining overall energy intake within a reasonable range. Because the LLM is constrained to operate only over foods retrieved from the FPED-based embedding space and to reference pre-computed HEI contributions, the resulting narratives remain aligned with the underlying nutrient data and HEI logic. Qualitative inspection of model outputs indicates that the HEI-aware RAG-LLM tends to recommend changes that are both nutritionally meaningful and behaviorally plausible. Typical substitutions include replacing high-sodium mixed dishes with lower-sodium alternatives in the same food category, swapping refined grains for whole-grain options, and adding fruits, vegetables, or legumes as side dishes. The framework also proposes portion-scaled variants to facilitate gradual shifts in intake. Further, as a basic feasibility check, most simulated profiles exhibited modest changes in estimated daily energy rather than extreme deviations, consistent with the utility function's design to penalize large departures from baseline intake. Although a comprehensive component-level analysis was beyond the scope of this initial study, the observed patterns align with the intended behavior of the HEI-aware retrieval mechanism, which prioritizes foods predicted to improve low-scoring dietary components.

Limitations and Future Work: Several data and modeling assumptions in the proposed framework can introduce potential inaccuracies. Nutrient and food composition databases differ in coverage and portion accuracy, which can propagate uncertainty into both HEI scoring and nutrient estimation. FPED mappings and HEI scoring rules, while validated for population-level surveillance, may not fully capture individual dietary needs or culturally specific dietary patterns. Candidate food sets may underrepresent regional, cultural, and taste preferences. In addition, LLMs, even when constrained via RAG, can over-generalize or generate recommendations that require tighter guardrails. Although mobile platforms have the potential to support real-time dietary guidance, most existing applications emphasize calorie tracking and self-monitoring rather than delivering personalized, evidence-based meal suggestions. Conventional diet mobile applications emphasize calorie counting and generic targets, shifting cognitive and data-entry implications to users without offering accurate and concrete diet guidance (Franco et al., 2016). Although we implemented the HEI-informed RAG-LLM framework within a prototype mobile application to demonstrate the feasibility of real-time food recommendations, a comprehensive evaluation of usability and real-world behavior change across diverse user groups with varying health contributions and goals remains an important direction for future work. Future research will advance this work by: (i) formally quantifying component-specific dietary changes and assessing their validity across key demographic and behavioral subgroups; (ii) expanding culturally diverse and preference-aware food databases to better capture heterogeneous eating patterns; (iii) enhancing LLM constraint mechanisms to ensure accuracy, safety, and personalization; and (iv) conducting rigorous assessments of usability, long-term adherence, and real-world dietary impacts through randomized controlled trials.

## 4 Conclusion

The study presented a HEI-informed RAG-based LLM framework for precision food recommendation. By anchoring retrieval in NHANES and FPED and scoring diet quality with the HEI, the approach advances toward recommendations that are interpretable and directly tied to established dietary guidelines. The proposed framework combines an embedding-based retrieval and scoring module that prioritizes foods expected to improve HEI, and a constrained RAG pipeline, instantiated with ChatGPT-class LLMs, that generates personalized suggestions with sourced guidelines. In our experiment on 12,076 entities, the proposed system achieved an average HEI increase of 6.45, with the proportion of users above HEI > 50 rising from 45.12% to 61.26%. Quantile analyses showed consistent rightward shifts in the HEI distribution, with the 25th, 50th, and 75th percentiles all increasing by more than 6. These results suggest that personalized AI-generated advice has the potential to meaningfully improve diet quality across a broad range of users. However, the evaluation is limited since it relies on simulated adoption of recommendations rather than observed behavior change. Further, HEI, while widely used, does not capture all dimensions of individual nutritional needs, preferences, or cultural context. The proposed framework may also produce biases and coverage limitations from NHANES, FPED, and current LLMs. Future work should therefore focus on prospective user studies to quantify real-world adherence and long-term changes in diet quality across clinical subgroups, and refinement of the retrieval and constraint mechanisms to better reflect condition-specific guidance. Extending the framework to alternative diet quality indices and incorporating dynamic feedback from ongoing intake data are additional opportunities.

**Acknowledgments**

This publication is supported by the Texas A&M AgriLife Institute for Advancing Health Through Agriculture (IHA) Research Capacity Funding and Texas A&M AgriLife Research. This research is partially supported by the United States Department of Agriculture (USDA)'s National Institute of Food and Agriculture (NIFA) Research Capacity Fund Hatch Program: TEX09954 (Accession No. 7002248). Any opinions, findings, conclusions, or recommendations expressed in this publication are those of the authors and should not be construed to represent any official USDA or U.S. Government determination or policy.

**Authorship contributions**

The authors' responsibilities were as follows – **YW:** Conceptualization, Methodology, Investigation, Writing – Original draft; **YY:** Conceptualization, Investigation, Writing – Original draft; **GG:** Conceptualization, Supervision, Writing – reviewing & editing; **RN:** Conceptualization, Supervision, Writing – reviewing & editing; **AZ:** Conceptualization, Methodology, Investigation, Supervision, Writing

– reviewing & editing, Resources, Funding acquisition. All authors read and approved the final manuscript.

**Data Availability**

The data supporting this work are derived from online repositories and previously published studies, which are cited in the references.